\documentclass[letterpaper, 10pt, conference]{ieeeconf}      
\IEEEoverridecommandlockouts                              % This command is only needed if 
\makeatletter
\let\NAT@parse\undefined
\makeatother

\usepackage[dvipsnames]{xcolor}

\usepackage{epsfig}
\usepackage{epstopdf}

\newcommand*\linkcolours{ForestGreen}

\usepackage{times}
\usepackage{graphicx}
\usepackage{amssymb}
\usepackage{gensymb}
\usepackage{amsmath}
\usepackage{breakurl}

\usepackage{url,hyperref}
\hypersetup{
colorlinks,
linkcolor=\linkcolours,
citecolor=\linkcolours,
filecolor=\linkcolours,
urlcolor=\linkcolours}

\usepackage{algorithmic}
\usepackage[vlined,ruled,linesnumbered]{algorithm2e}

\usepackage[labelfont={bf},font=small]{caption}
\usepackage[none]{hyphenat}

\usepackage{mathtools, cuted}

\usepackage[noadjust, nobreak]{cite}
\usepackage{tabularx}
\usepackage{amsmath,amssymb,color}

\usepackage{float}

\usepackage{pifont}% http://ctan.org/pkg/pifont
\newcommand{\subscr}[2]{{#1}_{\textup{#2}}}

\newcommand{\abs}[1]{|{#1}|}

\newcolumntype{Y}{>{\centering\arraybackslash}X}

\usepackage[]{placeins}

\usepackage{placeins}

\usepackage{tikz}

\usepackage[framemethod=tikz]{mdframed}

\usepackage{afterpage}

\usepackage{stfloats}

\usepackage{atbegshi}
\newcommand{\handlethispage}{}
\newcommand{\discardpagesfromhere}{\let\handlethispage\AtBeginShipoutDiscard}
\newcommand{\keeppagesfromhere}{\let\handlethispage\relax}
\AtBeginShipout{\handlethispage}

\usepackage{comment}

\title{\LARGE \bf
Competitive Perimeter Defense on a Line
}

\author{Shivam Bajaj$^{1}$ \qquad  Eric Torng$^{2}$ \qquad Shaunak D. Bopardikar$^{1}$
\thanks{$^{1}$Shivam Bajaj and Shaunak D. Bopardikar are with the Department of Electrical and Computer Engineering, Michigan State University. email:
    \texttt{bajajshi@msu.edu}. $^{2}$Eric Torng is with the Department of  Computer Science and Engineering, Michigan State University}
    \thanks{This work has been supported in part by NSF Award ECCS-2030556.}
}

\usepackage{theorem}
\newtheorem{lemma}{Lemma}
\newtheorem{definition}{Definition}

 \newtheorem{theorem}{Theorem}[section]
 
 \newtheorem{corollary}[theorem]{Corollary}

\newcommand{\E}{\mathcal{E}\xspace}

\begin{document}

\maketitle
\thispagestyle{empty}
\pagestyle{plain}

%%%%%%%%%%%%%%%%%%%%%%%%%%%%%%%%%%%%%%%%%%%%%%%%%%%%%%%%%%%%%%%%%%%%%%%%%%%%%%%%
\begin{abstract}
We consider a perimeter defense problem in which a single vehicle seeks to defend a compact region from intruders in a one-dimensional environment parameterized by the perimeter size and the intruder-to-vehicle speed ratio. The intruders move inward with fixed speed and direction to reach the perimeter. 
We provide both positive and negative \emph{worst-case} performance results over the parameter space using competitive analysis.
We first establish fundamental limits by identifying the most difficult parameter combinations that admit no $c$-competitive algorithms for any constant $c\geq 1$ and slightly easier parameter combinations in which every algorithm is at best $2$-competitive. We then design three classes of algorithms and prove they are 1, 2, and 4-competitive, respectively, for increasingly difficult parameter combinations.
Finally, we present numerical studies that provide insights into the performance of these algorithms against stochastically generated intruders. 
\end{abstract}
%%%%%%%%%%%%%%%%%%%%%%%%%%%%%%%%%%%%%%%%%%%%%%%%%%%%%%%%%%%%%%%%%%%%%%%%%%%%%%%%
\section{INTRODUCTION}
This paper addresses a perimeter defense problem, which is a class of vehicle routing problems, in which a vehicle seeks to intercept mobile intruders before they reach a specified region. In our problem, a robotic vehicle must defend a subset of a line segment from intruders that are generated at the endpoints of the line segment and move towards the subset, with a fixed speed. The robotic defender moves with maximum unit speed with the goal of \emph{capturing} the maximum number of intruders.
This perimeter defense problem is an \emph{online} problem in that the input, consisting of the arrival of intruders at specified times and locations, is only revealed gradually over time.

While perimeter defense problems have been well-studied, most prior work has focused on determining an optimal strategy for a small number of intruders or assuming that the input instance is generated by some stochastic process. While these results provide valuable insights into the average-case performance of defense strategies, they essentially ignore the worst-case where intruders may coordinate their actions to overwhelm the defense.

To understand how algorithms that specify vehicle motion perform in the worst-case, we adopt a competitive analysis technique~\cite{sleator1985amortized}. In competitive analysis, we measure the performance of an online algorithm $A$
% that plans the vehicle's movements
using the concept of \emph{competitive ratio}, i.e., the ratio of an optimal offline algorithm's performance divided by algorithm $A$'s performance for a worst-case input instance. Algorithm $A$ is $c$-competitive if its competitive ratio is no larger than $c$ which means its performance is guaranteed to be within a factor $c$ of the optimal for all input instances.

The primary application for our work is defending a perimeter from intruders such as missiles or locusts. Additional applications include gathering information on mobile entities in surveillance or traffic scenarios.\\

% \subsection{Related Work}
Perimeter defense problems were first introduced for a single vehicle and a single intruder in~\cite{isaacs1999}. Since then, perimeter defense has been mostly formulated as a pursuit-evasion game. The multiplayer setting for the same has been studied extensively as a reach-avoid game in which the aim is to design control policies for the intruders and the defenders \cite{chen2016multiplayer,  garcia2019strategies
% , fisac2015reach,yan2019task,
}. 
A typical approach requires computing solutions to the Hamilton-Jacobi-Bellman-Isaacs equation which is suitable for low dimensional state spaces and in simple environments \cite{margellos2011hamilton,chen2014path}. Recent works include \cite{yan2019matchingbased}, which proposes a receding horizon strategy based on maximum matching, and  \cite{ shishika2019perimeter}, which considers a scenario where the defenders are constrained to be on the perimeter. 

In vehicle routing problems, inputs
% may not be known in advance but 
become
available over time. Introduced on graphs in~\cite{psaraftis1988dynamic}, a typical approach requires that the vehicle routes be re-planned as new information is revealed over time
%becomes available over time bertsimas1993stochastic,
\cite{ bertsimas1991stochastic}.
The inputs may have multiple levels of priorities \cite{smith2010dynamic} or can be randomly recalled~\cite{bopardikar2020}. We refer the reader to \cite{bullo2011dynamic} for a review of this literature. 
A common way to analyze the performance of online algorithms is competitive analysis \cite{angelelli2007competitive, blom2001online,  henn2012algorithms}.
% which was introduced in \cite{sleator1985amortized}
% Competitive analysis has also been used in robotic navigation and exploration applications \cite{isler2003local,vander2015algorithms,ausiello2001algorithms}.
% \cite{de2002complexity} is a thesis on competitive analysis for vehicle routing algorithms.

Another area of related work is the class of Moving Target Travelling Salesman Problem (MTTSP) on a single line \cite{stieber2015multiple, hassoun2019single}. Several variants of this problem are discussed in \cite{helvig1998moving}.
Specifically, the authors provide an $\mathcal{O}(n^2)$ algorithm to capture $n$ intruders on a line in minimum time.
% A slightly similar work is in \cite{jindal2011multiple} wherein the authors consider that the intruders move away from an origin on a straight line. The authors consider two distinct objective functions, minimum total time and minimum total distance with resupply.\\

Earlier, we introduced a perimeter defense problem for a circular and rectangular environment with stochastically generated input~\cite{ShivamDVR2019, Smith2009translating}. The key distinction of the current work from these past works is the characterization of \emph{competitiveness} of the algorithms for worst case inputs.\\
\nocite{borodin2005online,krumke2003news}

% \subsection{Contributions and Organization}
Our general contribution is the following: we introduce a perimeter defense problem against mobile intruders using competitive analysis to derive worst-case performance guarantees. Specifically, we consider an environment comprising a line segment $[-1,1]$ in which the intruders arrive as per an arbitrary sequence at the endpoints. After arrival, the intruders move toward the origin, with fixed speed $v<1$, with the objective of reaching the region $[-\rho, \rho]$ called the \emph{perimeter}, for a given $0 < \rho < 1$. A vehicle, modelled as a first-order integrator with a maximum speed of unity, seeks to capture (become coincident with) the intruders before they reach the perimeter.  Our specific contributions are as follows.
%{\bf 1. Fundamental limits:} 
We first characterize a most difficult parameter regime in $v-\rho$ space in which no control algorithm (either online or offline) for the vehicle can be $c$-competitive for any constant $c$ and a second, slightly easier, parameter regime in which no algorithm is better than 2-competitive. 
We also show that a class of simple algorithms, such as the First-Come-First-Served, are not $c$-competitive, even for parameter regimes in which other algorithms are.
%{\bf 2. Algorithm design and analysis:} 
Next, we design and analyze three algorithms establishing $1$, $2$, and $4$-competitiveness, respectively, for increasingly difficult parameter regimes. 
% Specifically, we propose and analyze an open loop \emph{Sweeping} algorithm and characterize parameter regimes under which this algorithm is $1$-competitive, i.e., optimal. We propose and analyze a \emph{Compare and Capture} algorithm and characterize parameter regimes under which this algorithm is $2$-competitive. Finally, we propose and analyze a \emph{Capture with Patience} algorithm and characterize  parameter regimes for which it is $4$-competitive.
%{\bf 3. Numerical performance characterization:}
% While competitive analysis provides us with a worst-case performance guarantee on the performance of algorithms for certain parameter settings, it fails to provide us with a ``typical" performance on a random input with no coordination of intruders. To fill this gap, 
We numerically characterize their performance
% of these algorithms
when the intruders are generated as per a stochastic process.
% measuring algorithm performance by the percentage of intruders captured. 
We observe that the algorithms capture at least half the intruders generated even for parameter settings beyond their respective parameter regime.
% the Compare and Capture algorithm captures more than half the intruders generated even for parameter settings beyond its 2-competitive parameter regime and that and that the the number of intruders captured by the Sweeping algorithm converges to the number of intruders captured by Compare and Capture algorithm as $v$ increases. We further observe that the number of intruders captured by Capture with Patience algorithm is, on average, constant for a fixed $\rho$ and increasing $v$.

% \subsection{Organization}
% \sdb{[If we run short on space, we could merge this detail with the contributions paragraph above.]}
The paper is organized as follows. In Section \ref{sec:Problem}, we formally define our problem and the competitive ratio. We derive fundamental limits 
%in the parameter regime of $v$ and $\rho$ 
on how competitive any algorithm can be for difficult parameter regimes in Section \ref{sec:fundamental_limits}. In Section \ref{sec:algorithms}, we design and analyze three algorithms. Section \ref{sec:numerics} presents  numerical simulations. In Section \ref{sec:conclusion}, we present a summary and possible directions for future work. 

\section{Problem Formulation}\label{sec:Problem}
Consider an environment $\mathcal{E}(\rho)=\{y \in \mathbb{R} \, : \, \rho \leq \abs{y} \leq 1 \}$ and let $\bar{\mathcal{E}} := [-1,1]$. Intruders arrive over time at either location $-1$ or $1$ and move, with fixed speed $v<1$, towards the nearest point in $\bar{\mathcal{E}}$ out of  $-\rho$ or $\rho$.
The defense consists of a single vehicle with motion modeled as a first order integrator. The vehicle can move with a maximum speed of unity. The vehicle is said to \emph{capture} an intruder when the vehicle's location coincides with it. An intruder is said to be \emph{lost} if the intruder reaches the perimeter without being captured. Let $n(t)$ denote the number of intruders arrived at time instant $t$. An input instance $I$ is a tuple comprising the time instants, the corresponding number of intruders generated at those instants and their corresponding initial locations, is defined as $I = \{t,n(t),\{-1,1\}^{n(t)}\}_{t=0}^T$.

% An online algorithm does not have access to any arrival information for intruders until the moment they arrive at either location $-1$ or $1$.
% Thus, 
An online algorithm determines the velocity for the vehicle as a function of the location of the intruders that have arrived in the environment up to the current time instant $t$. Let $\mathcal{Q}(t,I)$ denote the set of instantaneous locations of all intruders in the environment at time $t$ from the input instance $I$. An intruder is removed from $\mathcal{Q}(t,I)$ if it is captured or lost.
We now formally define an online algorithm as follows.

\textit{Online Algorithm:} An online algorithm for a vehicle is a map $\mathcal{A} : \bar{\mathcal{E}}\times\mathbb{F} \to [-1,1]$, where $\mathbb{F}(\mathcal{E})$ is the set of finite subsets of $\mathcal{E}$, assigning a commanded velocity to the vehicle as a function of the position of the vehicle, denoted as $x(t)$, and the location of the intruders, yielding the kinematic model,
%\begin{align*}
    $\dot{x}(t) = \mathcal{A}(x(t),\mathcal{Q}(t,I)).$
%\end{align*}
% where $x(t)$ is the position of the vehicle in $\bar{\mathcal{E}}$.
% \sdb{In this work, an online algorithm is assumed to be memoryless and depends only the present state of the vehicle and the intruders.}

An optimal \emph{offline algorithm} is a non-causal algorithm having complete information of the input instance $I$ at any time $t\leq T$ and thus, the velocity of the vehicle is a function of current and future locations of the intruders.

\begin{definition}[Competitive Ratio]
Given an $\mathcal{E}(\rho)$, an input instance $I$ for $\mathcal{E}(\rho)$, and a given online algorithm $\mathcal{A}$, let $n_{\mathcal{A}}(I)$ denote the number of intruders captured by the vehicle when following algorithm $\mathcal{A}$ on input instance $I$. Let $OPT$ denote the optimal algorithm that maximizes the number of intruders captured out of input instance $I$. Then, the competitive ratio of $\mathcal{A}$ on $I$ is defined as $c_{\mathcal{A}}(I) = n_{OPT}(I)/n_{\mathcal{A}}(I) \ge 1$, and the competitive ratio of $\mathcal{A}$ for environment $\mathcal{E}$ is $c_{\mathcal{A}}(\mathcal{E}) = \sup_I c_{\mathcal{A}}(I)$. Finally, the competitive ratio for environment $\mathcal{E}$ is $c(\mathcal{E}) = \inf_{\mathcal{A}} c_{\mathcal{A}}(\mathcal{E})$.
We say that an algorithm is $c$-competitive for $\E$ if $c_{\mathcal{A}}(\E) \le c$.
\end{definition}

We assume that all of the input instances are non-adaptive where the arrival of intruders is not based on the movement of the vehicle.

\textit{Problem Statement:} The aim of this paper is to design $c$-competitive algorithms for the vehicle with minimum $c$.

\medskip
\section{Fundamental Limits}\label{sec:fundamental_limits}

We begin by characterizing a property of an \emph{extreme speed algorithm}, i.e., $\mathcal{A}' : \bar{\mathcal{E}}\times\mathbb{F} \to \{-1,0,1\}$, which either moves the vehicle with unit speed or keeps it stationary.

\begin{lemma}[Extreme speed algorithms]\label{lem:maxspeed}
For non-adaptive input instances where the arrival of intruders is not based on the movement of the vehicle, extreme speed algorithms are as powerful as general algorithms which can move with any speed in the range  $0\le V \le 1$. 
% \sdb{[How about a slight rephrase?]: Given an arbitrary algorithm $\mathcal{A}$, there exists an algorithm in which the speed of the vehicle is either maximum or zero at any given time, and that intercepts at least the same number of intruders as $\mathcal{A}$.}
\end{lemma}
\begin{proof}
% \sdb{One suggestion is something on these lines: Consider an instance of an arbitrary algorithm $\mathcal{A}$ and there is an interval of time $\Delta T$ over which the vehicle's speed $\abs{\dot{x}(t)} < 1$, and let us assume that in this interval $\Delta T$, the vehicle's direction does not change. Then, consider a second algorithm that moves the vehicle with maximum speed in the same direction as $\mathcal{A}$, stops and then reverses its direction after a time interval of $\Delta T_1 < \Delta T$ so that at the end of the interval $\Delta T$, the vehicle becomes coincident with its location while following $\mathcal{A}$. The distance covered by the vehicle in the latter case exceeds the former and therefore, the number of intruders intercepted in the interval $\Delta T$ is guaranteed to be at least the same as that for the arbitrary algorithm. }
% Clearly, a general algorithm $\mathcal{A}$ that moves the vehicle with any speed is at least as powerful as an extreme speed algorithm $\mathcal{A'}$ that can only move the vehicle with speed 1 or keep the vehicle stationary.
% We now prove the reverse. \sdb{[I think we can skip the first paragraph, given the space limitation and begin with what is below.]}
Let $\mathcal{A}$ denote an arbitrary general speed algorithm and $I$ denote an arbitrary non-adaptive input instance where there are $n$ distinct time points where $\mathcal{A}$ captures at least one intruder of $I$.
We define the \emph{capture profile} of $\mathcal{A}$'s execution on $I$ as
%\textcolor{ForestGreen}{We observe that any intruder $j$ in $I$ that is captured by $\mathcal{A}$ can be associated with a pair $(x_j,k_j)$, where $x_j$ and $k_j$ is the capture location and time when intruder $j$ is captured.
%We denote the \textit{fingerprint} of $\mathcal{A}$ as}
the set of pairs $(x_i,k_i)$ for $1\le i \le n$ where $k_i$ is the time of the $i$th capture and $x_i$ is the point where the $i$th capture occurred.
Let $I(x_i,k_i)$ denote the intruders in $I$ captured by $\mathcal{A}$ at location $x_i$ at time $k_i$.
We add to this capture profile the pair $(x_0,k_0)$ which denotes the starting location for the vehicle and the starting time instant.

Our goal is to show that there exists an extreme speed algorithm $\mathcal{A'}$ that can match $\mathcal{A}$'s capture profile for $I$: namely having the vehicle at location $x_i$ at time $k_i$ for $0 \le i \le n$. If we can show this, then because $I$ is non-adaptive, $\mathcal{A'}$ will capture the intruders in $I(x_i,k_i)$ for $1\le i \le n$  at $x_i$ at time $k_i$ and the result follows.
Note that $\mathcal{A'}$ may capture the intruders in $I(x_i,k_i)$ at an earlier time than $k_i$ because $\mathcal{A'}$ may move differently than $\mathcal{A}$; the  crucial point is that $\mathcal{A'}$ captures every intruder captured by $\mathcal{A}$.

We now show that there exists an $\mathcal{A'}$ that can match $\mathcal{A}$'s capture profile for $I$.
We prove this by induction on $i$; namely, that there exists an $\mathcal{A'}$ that will have the vehicle at point $x_j$ at exactly time $k_j$ for $0 \le j \le i$ for $0 \le i \le n$.
The base case where $i=0$ is trivial since the vehicle starts at location $x_0$ at time $k_0$ for both algorithms.
We now show that if this is true up to $i$, then it also holds for $i+1$.
Given the induction hypothesis, we know that there exists an $\mathcal{A'}$ that will have the vehicle at point $x_j$ at exactly time $k_j$ for $0 \le j \le i$.
We extend that $\mathcal{A'}$ by showing it can also have the vehicle at $x_{i+1}$ at time $k_{i+1}$.

%Consider an algorithm $\mathcal{A}$ that moves the
%vehicle, with any speed $V\leq1$, from $x_i$ for $0\leq i\leq n$ at time instant $k_i$ to the next location $x_{i+1}$ arriving at time instant $k_{i+1}$, where $n$ denotes the total number of distinct capture locations. The location $x_i$ and time instant $k_i$ denote the location and time instant when the vehicle captures intruders with an exception that $x_0$ and $k_0$ denotes the starting location and time instant of the vehicle. Basically, $k_{i+1}-k_i$ is the time interval between two successive capture of the intruders by the vehicle.
%Without loss of generality, we assume $x_0$ and $k_0$ to be $0$. Consider another algorithm $\mathcal{A}'$, in which the vehicle starts at the same initial location as algorithm $\mathcal{A}$, which moves the vehicle with maximum speed or keeps the vehicle stationary. The idea is to show that algorithm $\mathcal{A}'$ can have the vehicle located at the same location and same time instant as algorithm $\mathcal{A}$ in the start and end of each interval. This is because, the starting and ending location and time instant of each interval is the location and time instant when the vehicle captures the intruders. If the vehicle, according to algorithm $\mathcal{A}'$, is at the same starting and ending location and time instant as algorithm $\mathcal{A}$, then it means that the vehicle will capture the same intruders at the same time and location as algorithm $\mathcal{A}$ and thus, achieve the same performance. 

At time $k_i$, we have $\mathcal{A'}$ move the vehicle from $x_i$ to $x_{i+1}$ arriving at $x_{i+1}$ no later than $k_{i+1}$ since the the maximum speed of $\mathcal{A}$'s vehicle is 1, just like $\mathcal{A'}$'s vehicle. We then have $\mathcal{A'}$ keep the vehicle at $x_{i+1}$ until time $k_{i+1}$.
This completes the inductive step and, in turn, completes the proof.
\end{proof}

In light of Lemma~\ref{lem:maxspeed}, we can restrict our attention to algorithms that either move the vehicle with maximum speed or keep the vehicle at rest.

% [NOTE: I am thinking about the utility of this result -- the only place where it arises is in the analysis of optimal algorithms. Perhaps, we might want to revisit that analysis to see if Lemma 1 is indeed required or not.]}

We now present the fundamental limits.
We first present a fundamental limit on achieving a $c$-competitive ratio for any constant $c$ followed by a fundamental limit on achieving a $2$-competitive ratio.
\begin{theorem}\label{thm:Inc_no_c}
For any environment $\mathcal{E}$ such that $v>\frac{1-\rho}{2\rho}$,
\renewcommand{\theenumi}{\roman{enumi}} 
\begin{enumerate}
    \item there does not exist a $c$-competitive algorithm and
    \item no algorithm (online/offline) can capture all intruders.
\end{enumerate}

\end{theorem}
\begin{proof}
% The idea behind the proof is to first construct a worst case input for any online algorithm and then compare the performance with the optimal algorithm. Note that the optimal algorithm has access to offline/future information of the entire input instance. \\ 
Assume that the vehicle starts at the origin. The input instance consists of two phases: a stream of intruders that arrive at the endpoint $1$, $2$ time units apart starting at time $1$, and a burst of $c+1$ intruders who arrive at the endpoint $-1$ at time $t$ that corresponds to the first time the vehicle moves to $\rho$ according to any online algorithm.
Note that if the vehicle never moves to location $\rho$, the stream never ends, and the burst never arrives, so the algorithm will not be $c$-competitive for any constant $c$, and the first result follows.
Let $i$ be the number of stream intruders released up to and including time $t$; note that $i$ might be 0 if the vehicle reaches $\rho$ before time 1 when the first stream intruder is released. 
We first observe that the vehicle will capture at most intruder $i$. In particular, because the stream intruders are released $2$ time units apart and $v > \frac{1-\rho}{2\rho}$, all stream intruders before $i$ have reached $\rho$ before time $t$. Two cases arise: $i=0$ and $i > 0$.\\
Case  1: If $i=0$, this means the vehicle reached location $\rho$ before time $1$, so $t<1$. Since $v>\frac{1-\rho}{2\rho}$, the vehicle will not be able to capture the burst of $c+1$ intruders that arrive at time $t$. The optimal offline algorithm, however, can move the vehicle to location $-\rho$ by time $t$ and will thus capture all $c+1$ burst intruders.\\
Case 2: If $i >0$, this means the vehicle reached location $\rho$ no earlier than time 1, so $1 \le t$.
In this case, the optimal algorithm can capture the first $i-1$ stream intruders immediately upon arrival by moving the vehicle to endpoint $1$ at time $1$ and staying there until the first $i-1$ stream intruders have been captured. The algorithm then moves the vehicle to $-\rho$ before the burst intruders have been released since the stream intruders arrive $2$ time units apart. If $i=1$, the algorithm moves the vehicle immediately to $-\rho$.\\
In either case, we show that the optimal offline algorithm can capture all the burst intruders and at least all but the last stream intruder whereas the online algorithm will capture at most 1 stream intruder, and the first result follows.\\
The second result follows by observing there are choices for $t$ including $t=1$ such that no algorithm can capture all $i$ stream intruders and all $c+1$ burst intruders.
\end{proof}

% Theorem \ref{thm:Inc_no_c} provides a parameter regime in which no $c$-competitive algorithm exists.
The following theorem provides a fundamental limit on achieving a $2$-competitive ratio for any environment.

\begin{theorem}\label{lem:bound_2_comp_IncP}
For the environment $\E$ such that $v \ge \frac{1-\rho}{1+\rho}$, $c(\E) \ge 2$.
\end{theorem}
\begin{proof}
In this proof, all of our input instances consist of two intruders, $a$ and $b$, where $a$ arrives at endpoint $-1$ and $b$ arrives at endpoint $1$ and we assume that the vehicle starts at the origin at time $0$.
We first consider the case where $v = \frac{1-\rho}{1+\rho}$.
Consider the following input instance $I_1$ where both $a$ and $b$ arrive at time $1$.
For input instance $I_1$, there are two symmetric solutions to capture both intruders; we describe one below.
At time $0$, the vehicle moves toward endpoint $1$ capturing intruder $b$ at endpoint $1$ at exactly time $1$.
The vehicle then moves to $-\rho$ to capture intruder $a$ at $-\rho$ at exactly time $2+\rho$. The vehicle has just enough time to do this given the condition that $v = \frac{1-\rho}{1+\rho}$.
Thus, any algorithm that hopes to be better than $2$-competitive for such an $\E$ must capture both intruders in $I_1$, and the only way to do so is to move immediately to $-1$ or $1$ arriving at the destination point at exactly time $1$.

Now consider input instances $I_2$ and $I_3$ which also consists of two intruders. In $I_2$, intruder $a$ arrives at time $1$ and intruder $b$ arrives at time $1+\epsilon$ where $\epsilon < 2 \rho v$. In $I_3$, intruder $b$ arrives at time 1 and intruder $a$ arrives at time $1+ \epsilon$ for same $\epsilon$.
Algorithms which have the vehicle arriving at endpoint $1$ at time $1$ can capture at most one of the two intruders in input instance $I_2$. This follows since the vehicle can only capture intruder $a$ if it moves immediately to arrive at $-\rho$ at time $2+\rho$. However, as $\epsilon < 2 \rho v$ and $v = \frac{1-\rho}{1+\rho}$, the vehicle cannot capture intruder $b$ before $b$ passes $\rho$.
Similarly, algorithms which have the vehicle arrive at $-1$ at time $1$ can capture at most one of the two intruders in input instance $I_3$.
At the same time, there exists an offline algorithm which captures both intruders in input instance $I_2$ and another algorithm which captures both intruders in $I_3$ which is to simply move to the correct location at time $1$ and then capture the other intruder before it reaches $\rho$ or $-\rho$.\\
We now consider the case where $v > \frac{1-\rho}{1+\rho}$.
Now we only need two input instances $I_4$ and $I_5$.
In $I_4$, intruder $a$ arrives at time $1$ and intruder $b$ arrives at time $1+\epsilon$ where $\epsilon = 1+\rho - \frac{1-\rho}{v}$. In $I_5$, intruder $b$ arrives at time $1$ and intruder $a$ arrives at time $1+ \epsilon$ for same $\epsilon$.
As $v > \frac{1-\rho}{1+\rho}$, $\epsilon > 0$,
using similar arguments as for $I_2$ and $I_3$, it follows that no single online algorithm can capture both intruders in $I_4$ and $I_5$.
%but there exist different algorithms which can capture both points in $I_4$ and $I_5$, so the result follows.

In summary,
even restricting the set of possible input instances to $\{I_1, I_2, I_3,I_4,I_5\}$, no single algorithm can capture both intruders from all five input instances, but since there exist algorithms which capture both intruders for all five input instances, it follows that $c(\E) \ge 2$.
\end{proof}

We now show that a natural algorithm, \emph{First-Come-First-Served} (FCFS), is not an effective algorithm for this problem.
We define FCFS as the algorithm which sends the vehicle at speed $1$ towards the earliest intruder to arrive that is not lost
% (guaranteed to reach its goal before the vehicle)
or already captured breaking ties arbitrarily.

\begin{lemma}
For any $\E$ where $\frac{2}{v+1}+\rho > \frac{1-\rho}{v} + \epsilon$ for some small $\epsilon > 0$, FCFS is not $c$-competitive for any constant $c$.
\end{lemma}
\begin{proof}
We prove this result by constructing a worst case input instance against FCFS. Consider the following input instance $I(c)$. 
Let the first intruder be released at time $0$ at $1$.
Let $c+1$ intruders be released at time $\epsilon$ at $-1$.
FCFS will move the vehicle immediately towards the first intruder and intercept it at $\frac{1}{v+1}$ at time $\frac{1}{v+1}$.
Because of the condition $\frac{2}{v+1}+\rho > \frac{1-\rho}{v} + \epsilon$, FCFS cannot get the vehicle to $-\rho$ before the $c+1$ intruders released at time $\epsilon$ reach $-\rho$.
It follows that $FCFS(I(c)) = 1$.
On the other hand, if the vehicle had moved toward $-\rho$ immediately, it would capture those $c+1$ intruders.
Thus, $OPT(I(c)) = c+1$, and the result follows. This result generalizes to any variation of FCFS which services the first arriving intruder, if possible, before any later arriving intruder.
\end{proof}

We now turn our attention to the design of algorithms with provable guarantees on the competitive ratio.
In the following section, we describe and analyze three algorithms, characterizing parameter regimes with provably finite competitive ratios.

\section{Algorithms}\label{sec:algorithms}
We now propose three main algorithms for the vehicle that are provably $1$, $2$, and $4$ competitive. As the competitive ratio increases, the parameter regime that can be handled also increases.

\subsection{Sweeping algorithm}
We define the Sweeping algorithm (Sweep) as follows.
%In this algorithm, 
At time 0, the vehicle moves with unit speed toward endpoint $+1$. %, i.e., either $+1$ or $-1$. 
From this point on, the vehicle only changes direction when it reaches an endpoint $+1$ or $-1$, at which time it moves with unit speed towards the opposite endpoint.
Sweep is an \emph{open-loop} algorithm; that is, it ignores all information about intruders. 
One logical variant is to stop moving in a given direction if there are no intruders in that direction. We show this variant achieves the same performance guarantee.

\begin{theorem}\label{lem:1_comp}
For environment $\mathcal{E}$, Sweep is 1-competitive if 
$v \leq \frac{1-\rho}{3+\rho}$.
If not, it is not $c$-competitive for any constant $c$.
\end{theorem}
\begin{proof}
Suppose that $v \leq \frac{1-\rho}{3+\rho}$ holds.
We show Sweep captures all intruders.
Any intruder $i$ will take $\frac{1-\rho}{v}$ time to get from its arrival location, which we now assume to be $1$ without loss of generality, to $\rho$.
In the worst case for Sweep, which is that it has left $1$ just before intruder $i$ arrived, it will take $3+\rho$ time to get to $\rho$ moving towards intruder $i$.
If $3+\rho \le \frac{1-\rho}{v}$, then Sweep's vehicle will get to $\rho$ first and intruder $i$ will be captured, and the first result follows.

For $v > \frac{1-\rho}{3+\rho}$ there is an input instance where intruders only arrive at $1$ just after the vehicle has left $1$.
As $v > \frac{1-\rho}{3+\rho}$, all intruders will be lost and the second result follows.
\end{proof}
We observe that the upper bound in the proof holds for the Sweep variant where it stops moving in a given direction if there are no intruders in that direction. The lower bound requires introducing some intruders at $-1$ to ensure that the modified Sweep will move the vehicle towards $-1$. These intruders might be captured, but the lower bound still holds by increasing the number of intruders which arrive at $1$.

\subsection{Compare and Capture (CaC) algorithm}
We now present a  Compare and Capture (CaC) algorithm that is provably $2$-competitive beyond the parameter regime of the Sweep algorithm.
CaC is not open-loop but is \emph{memoryless},
i.e., its actions depend only on the present state of the vehicle and the intruders. 

%In this subsection, we present a  algorithm and identify the range of parameters $v$ and $\rho$ for which it is 2-competitive. 
We begin with some notation and definitions. %, and then we define the algorithm. % followed by the algorithm and then our proof.
%Then we %will 
%characterize sufficient conditions on $v$ and $\rho$ such that CaC to be 2-competitive.\\
%
% We say that a set of intruders $S$ is on the $\emph{same side}$ as the vehicle if the vehicle is located at $\rho$ (resp.~$-\rho$) and $S \subset (\rho,1]$(resp. $[-1,-\rho)$). Similarly, we say that $S$ is on the $\emph{opposite side}$ of the vehicle if the vehicle is located at $\rho$(resp.~$-\rho$) and $S \subset [-1,-\rho]$ (resp. $[\rho,1]$).\\
%
An epoch $k$ for the CaC algorithm is the time interval when the vehicle moves from location $x_k$ to location $x_{k+1}$ and is about to move from $x_{k+1}$, capturing some intruders along the way.
Location $x_k$ is always either $\rho$ or $-\rho$.
We denote the start of epoch $k$ using the notation $k_S$.
%, in order to capture some intruders, and before ending the epoch at some location $x_{k+1}$.
%Without loss of generality, let the vehicle be located at $x_k = \rho$ at the beginning of the $k$-th epoch. 
For epoch $k$, we define $\subscr{S}{same}^k$ as the set of intruders on the $\emph{same side}$ as the vehicle at time $k_S$, and $\subscr{S}{opp}^k$ as the set of intruders on the $\emph{opposite side}$ of the vehicle that are between $\rho+2\rho v$ and $\rho+2v\rho+\frac{2v(1-\rho)}{1+v}$  away from the origin at time $k_S$. Specifically, if the vehicle is located at $x_k=\rho$, then $\subscr{S}{opp}^k$ is defined as the set of intruders contained in $[-(\rho+2\rho v+\frac{2v(1-\rho)}{1+v}),-(\rho+2\rho v)]$.
%The $2\rho v$ term in the lower bound $\rho+2\rho v$ guarantees that if the vehicle moves from $x_k$ to $-x_k$, the vehicle will reach $-x_k$ before any intruder does. 
% See Figure~\ref{fig:proof} (a) for an illustration of 
% $\subscr{S}{same}^k$ and $\subscr{S}{opp}^k$ for a case where $x_k = \rho$.

%and are contained in $[-\rho-2v\rho-\frac{2v(1-\rho)}{1+v},-\rho-2\rho v]$ at $k^{th}$ epoch (Fig. \ref{fig:proof} (a)). Similarly, if the vehicle is located $x_k = -\rho$, $\subscr{S}{opp}^k$ is the set of points on the $\emph{opposite side}$ of the vehicle and are contained in $[\rho+2\rho v,\rho+2\rho v+\frac{2v(1-\rho)}{1+v}]$.

\medskip
The CaC algorithm, summarized in Algorithm~\ref{algo:CaC}, works as follows: At epoch $k$, for any $k\ge 1$, %let us assume that the vehicle is located at $x_k = \rho$. 
the algorithm computes the number of intruders located in $\subscr{S}{same}^k$ and $\subscr{S}{opp}^k$. If the total number of intruders in $\subscr{S}{same}^k$ is greater than the total number of intruders in $\subscr{S}{opp}^k$, then the vehicle moves away from the origin for at most $\frac{1-\rho}{1+v}$ time to capture all intruders from the set $\subscr{S}{same}^k$ and then returns to $x_{k+1} = x_k$.
% Then the vehicle waits for $\frac{2(1-\rho)}{1+v}-\subscr{T}{capt}^k$ time units at $x_{k+1}$, where $\subscr{T}{capt}^k$ denotes the total time taken by the vehicle to capture intruders in an epoch $k$.
Otherwise, the vehicle moves for at most $2\rho + \frac{4v(1-\rho)}{(1+v)^2}$ time to capture all intruders located in $\subscr{S}{opp}^k$ and then returns to $x_{k+1} = -x_k$.
% The vehicle waits for $\frac{1-\rho-2\rho v}{v}-\subscr{T}{capt}^k$ time units at $x_{k+1}$.
%  \erictodo{I now realize we don't need to worry about considering intruders twice. The key is there is a last set every intruder belongs to before it is either captured or lost. That is the one we will account for. This argument needs to move into the final proof of CaC being 2-competitive.}
% Note that, while comparing the number of intruders in $\subscr{S}{same}^{k+1}$ and $\subscr{S}{opp}^{k+1}$, the vehicle ignores the intruders already considered in epoch $k$. 

At time $0$, we assume the vehicle starts at the origin.
CaC waits at the origin until the first intruder that arrives in the environment is $3\rho v+\rho$ distance away from the origin, i.e., the vehicle does not move until $\frac{1-\rho -3\rho v}{v}$ time units after the first intruder arrives. If the total number of intruders located in $[\rho+3\rho v,1]$ is greater than the total number of intruders located in $[-1,-(\rho+3\rho v)]$, then the vehicle moves to $x_1=\rho$. If not, the vehicle moves to $x_1=-\rho$. The first epoch begins when the vehicle reaches $x_1$.

To prove $2$-competitiveness of CaC, we first prove that any intruder not belonging to $\subscr{S}{same}^k$ or $\subscr{S}{opp}^k$ in an epoch $k$ will not be lost during epoch $k$.
%We then prove that $\subscr{S}{opp}^k$ is well-defined \sdb{[Need to explain what ``well-defined" means here]}.
% ; in particular that $\rho+2\rho v+\frac{2v(1-\rho)}{1+v} \leq 1$.

% \begin{figure}[t]
%     \centering
%     \includegraphics[scale=0.7]{proof_1 - Copy.png}
%     \caption{ Illustration of $\subscr{S}{opp}^k$ and $\subscr{S}{same}^k$ by the vehicle in epoch $k$.
%     % Snapshot of epoch $k$. The dark red circles are the intruders considered in epoch $k$ and the bright red circles are the intruders considered in epoch $k+1$. (a) The vehicle compares $|\subscr{S}{same}^k|$ and $|\subscr{S}{opp}^k|$. (b) The vehicle moves on the $\emph{same side}$ to capture $|\subscr{S}{same}^k|$. (c) The vehicle returns to $\rho$ after capturing all intruders in $\subscr{S}{same}^k$. Intruders that did not belong to $\subscr{S}{same}^k$ and $\subscr{S}{opp}^k$ are not lost
%     }
%     \label{fig:proof}
% \end{figure}

\begin{algorithm}[t]
	\DontPrintSemicolon
	\SetAlgoLined
	Vehicle is at center and waits for $\frac{1-\rho-3\rho v}{v}$ time units\\
	\eIf{ intruders in $[\rho+3\rho v,1]\leq \text{ intruders in }[-1,-(\rho+3\rho v)]$}{
		Move to $-\rho$ 
		\;}
	{ Move to $\rho$.	\;
	}
%	At epoch $k$, assumes vehicle is located at $x_k = \rho$ \\
	\For{each epoch $k\geq1$}{
	\eIf{at epoch $k$, $|\subscr{S}{same}^k|\leq|\subscr{S}{opp}^k|$}{
		Move to capture all intruders located in $\subscr{S}{opp}^k$ \\
		Move to $x_{k+1} = -x_k$
% 		Wait for $\frac{2(1-\rho)}{v}-\subscr{T}{capt}^k$ time units
		\;}
	{ Move to capture intruders located in $\subscr{S}{same}^k$.\\
	Move to $x_{k+1} = x_k$
% 	Wait for $\frac{1-\rho-2\rho v}{v}-\subscr{T}{capt}^k$ time units.
	\;
	}}
	\caption{Compare-and-Capture Algorithm}
	\label{algo:CaC}
\end{algorithm}

% We will now provide a sufficient condition on the speed $v$ and $\rho$ which guarantees that at every epoch $k$, the vehicle can capture intruders that lie in $\subscr{S}{same}^k$ or $\subscr{S}{opp}^k$.

\begin{lemma}\label{lem:condition}
In every epoch $k$, any intruder that lies outside of the set $\subscr{S}{same}^k$ and $\subscr{S}{opp}^k$ is not lost if
%\begin{equation}\label{eq:compare}
    $\frac{\rho v}{1-\rho}+\frac{v^2}{(1+v)^2}\leq\frac{1}{4}.$
%\end{equation}
\end{lemma}
\begin{proof} Assume that the vehicle is located at $x_k=\rho$ at time $k_S$. Two cases arise:

\emph{Case 1: $|\subscr{S}{same}^k|>|\subscr{S}{opp}^k|$:}
In this case, the vehicle moves away from the origin to capture intruders located in $\subscr{S}{same}^k$. The vehicle takes at most $\frac{1-\rho}{1+v}$ time to capture these intruders. The total time taken by the vehicle to capture the intruders and return to $x_{k+1}=x_k$ in epoch $k$ is at most $\frac{2(1-\rho)}{1+v}$. Since the intruders in $\subscr{S}{opp}^k$ are located in $[-(\rho+2\rho v+\frac{2v(1-\rho)}{1+v}),-(\rho+2\rho v)]$, any intruder in the opposite side which did not belong to $\subscr{S}{opp}^k$ will be at least $2\rho v$ distance away from $\rho$ and thus, will be contained in $\subscr{S}{opp}^{k+1}$.

\emph{Case 2: $|\subscr{S}{same}^k|\leq|\subscr{S}{opp}^k|$:} The total time taken by the vehicle to capture intruders located in $\subscr{S}{opp}^k$ and return back to $x_{k+1}=-x_k$ is at most $2\rho + \frac{4v(1-\rho)}{(1+v)^2}$. In order to ensure that any intruder that did not belong to $\subscr{S}{same}^k$ is at least $\rho+2\rho v$ distance away from the origin at the end of epoch $k$, we require
%\begin{align*}
    $2\rho + \frac{4v(1-\rho)}{(1+v)^2} \leq \frac{1-\rho-2\rho v}{v}
    \Rightarrow \frac{\rho v}{1-\rho}+\frac{v^2}{(1+v)^2} \leq \frac{1}{4}.$
%\end{align*}
% Then, the total time of epoch $k$, i.e., time taken by the vehicle to capture intruders located in $\subscr{S}{opp}^k$ and return back to $-\rho$ is $2\rho + \frac{4v(1-\rho)}{(1+v)^2}$.
% In order to ensure that $\subscr{S}{opp}^{k+1}$ is completely capturable, it is sufficient to ensure that all intruders except the intruders located in $\subscr{S}{same}^{k+1}$ are at a distance of $\rho + 2\rho v$ from the origin (Fig. \ref{fig:proof} (c)). Mathematically,
% \begin{align*}
%     2\rho + \frac{4v(1-\rho)}{(1+v)^2} \leq \frac{1-\rho-2\rho v}{v}
%     \Rightarrow \frac{\rho v}{1-\rho}+\frac{v^2}{(1+v)^2} \leq \frac{1}{4}.
% \end{align*}
This concludes the proof.
\end{proof}

\begin{lemma}\label{lem:set_def_CaC_Inc}
In every epoch $k$ of the CaC, $\subscr{S}{opp}^k$ is well defined if
%\begin{equation}\label{eq:feasiblity}
    $\rho+2\rho v+\frac{2v(1-\rho)}{1+v} \leq 1.$
%\end{equation}
\end{lemma}
\begin{proof}
In order to ensure that $\subscr{S}{opp}^k$ is well defined, we require that $\subscr{S}{opp}^k$ is contained in the environment. Mathematically, this corresponds to
% \begin{align*}
    $\rho+2\rho v+\frac{2v(1-\rho)}{1+v} \leq 1$,
% \end{align*}
and the result follows. 
\end{proof}

\begin{theorem}\label{thm:2_comp}
Algorithm \ref{algo:CaC} captures at least half of all intruders and is 2-competitive in parameter regimes for which Lemma \ref{lem:condition} and Lemma \ref{lem:set_def_CaC_Inc} both hold.
%that satisfies equations \eqref{eq:compare} and \eqref{eq:feasiblity}.
\end{theorem}
\begin{proof}
% \erictodo{Based on the second lemma, $S_opp$ is well-defined so CaC is well-defined. Based on the first lemma, every intruder is part of some set $S$. For each intruder, associate it with the final set to contain that intruder. It is either lost or captured. CaC captures the entire set of the set it tries to get. Thus the captured set of each epoch is larger than the lost set of each epoch. Since every intruder is considered, the result follows.}
Based on Lemma \ref{lem:set_def_CaC_Inc}, $\subscr{S}{opp}^k$ is well-defined in every epoch $k$. This implies that Algorithm \ref{algo:CaC} is well-defined. Lemma \ref{lem:condition} ensures that every intruder will belong to  $\subscr{S}{same}^k$ or $\subscr{S}{opp}^k$ for some epoch $k$. 
%The waiting at $x_{k+1}$ ensures that all the intruders that vehicle decided to lose are lost such that they are not considered in the next epoch. Thus, any intruder that belongs to the set $\subscr{S}{same}^k$ or $\subscr{S}{opp}^k$ is either captured or lost.
Because of the comparison in line 9 of Algorithm \ref{algo:CaC}, in every epoch, the number of intruders captured is at least the number of intruders lost. Note that an intruder is not lost in epoch $k$ if it belongs to $\subscr{S}{same}^{k+1}$ or $\subscr{S}{opp}^{k+1}$  in the subsequent epoch $k+1$.
Thus, the algorithm captures at least half of all intruders and is $2$-competitive.
\end{proof}

\subsection{Capture with Patience (CAP) Algorithm}
We now present another memoryless algorithm, Capture with Patience (CAP), in which the vehicle stays in the range $[-\rho, \rho]$ waiting for and capturing intruders at $\rho$ or $-\rho$. CAP is only $4$-competitive but can operate beyond the parameter regime of the CaC algorithm. 

A key feature in environment $\E$ is the quantity $z = \frac{1-\rho}{v}$ which represents the time required for any intruder originating at $+1$ or $-1$ to reach the corresponding location $\rho$ or $-\rho$.
We first describe an algorithm that is provably $4$-competitive for $\E$ when $z \ge 6 \rho$, equivalently $v \le \frac{1-\rho}{6\rho}$. This requirement ensures that incoming intruders require at least $6\rho$ time to get to either $-\rho$ or $\rho$ whereas it takes the vehicle $2\rho$ time to move from $-\rho$ to $\rho$ or vice versa.

\begin{figure}[t]
    \centering
    \includegraphics[scale=0.5]{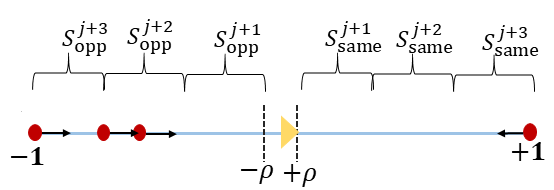}
    \caption{Breakdown of the environment into regions of length $2\rho v$ by the CAP algorithm. The triangle depicts the vehicle, whereas a dot depicts an intruder}
    \label{fig:CAP_region}
\end{figure}

The algorithm is formally defined in Algorithm \ref{algo:CAP_IncP} and is described as follows.
First, to simplify notation, it defines time 0 to be the moment when the first intruder arrives.
It then breaks time up into intervals of length $2\rho$. The $i$th interval for $i \ge 1$ is defined as the time interval $[2(i-1)\rho, 2i\rho]$. We say that a set of intruders $S$ is on the $\emph{same side}$ as the vehicle if the vehicle is located at $\rho$ (resp.~$-\rho$) and $S \subset (\rho,1]$ (resp. $[-1,-\rho)$). Similarly, we say that $S$ is on the $\emph{opposite side}$ of the vehicle if the vehicle is located at $\rho$ (resp.~$-\rho$) and $S \subset [-1,-\rho]$ (resp. $[\rho,1]$).\\
For $i \ge 1$, let $\subscr{S}{opp}^i$ and $\subscr{S}{same}^i$ be the intruders in an input instance $I$ that arrive in the $i$th time interval that are on the opposite side and same side of the vehicle, respectively. %, \sdb{and which are generated in} the $i$th time interval. 
Let $|S|$ denote the cardinality of $S$.
% let $\abs {\subscr{S}{L}^i}$ and $\abs{\subscr{S}{R}^i}$ denote the number of intruders in $\subscr{S}{L}^i$ and $\subscr{S}{R}^i$, respectively.

% For $i \ge 1$, let $L(I,i)$ and $R(I,i)$ be the points in $I$ that arrive at $-1$ or $1$, respectively, during the $i$th time interval, and let
% let $|L(I,i)|$ and $|R(I,i)|$ denote the number of points in $L(I,i)$ and $R(I,i)$, respectively.

% \erictodo{I modified this; I think steady state should start at time $z$ rather than $2\rho + z$}.
The algorithm  operates as follows in the steady state, i.e., after time instant $z$.
At any time instant $2j \rho+z$ for $j \ge 0$, 
%for $j \ge 1$,
the vehicle is stationed at either $-\rho$ or $\rho$.
Without loss of generality, we assume that the vehicle is stationed at $\rho$.
% Either, the vehicle has just moved here $-\rho$ to ensure it captures all the points in $R(I,j+1)$, or the vehicle captured all the points in $R(I,j)$ and must now decide whether to capture all the points in $R(I,j+1)$ or move to $-\rho$.
% For this latter case, it makes the decision as follows.
% First, we observe that the points in $R(I,j+1)$ are located between points $\rho$ and $\rho + 2\rho v$, the points in $R(I,j+2)$ are located between points $\rho + 2\rho v$ and $\rho+4\rho v$, and the points in $R(I,j+3)$ are located between points $\rho + 6\rho v$.
% Further, because $z \ge 6\rho$, this means all the points in $R(I,j+3)$ have arrived by time $2j \rho + z$. Similar conclusions can be drawn for the points in $L(I,j+1)$, $L(I,j+2)$, and $L(I,j+3)$. 
% If $|L(I,j+2)| > |R(I,j+1)| + |R(I,j+2)| + |R(I,j+3)|$, then the vehicle moves to $-\rho$ arriving at time $2(j+1)\rho + z$ which is just in time to capture all the points in $L(I,j+2)$. 
% If not, the vehicle stays at $\rho$ and captures all the points in $R(I,j+1)$ and then reevaluates at time $2(j+1)\rho +z$.
% The key observation is that the vehicle changes position only when it sees enough benefit in the points in $L(I,j+2)$ such that it can sacrifice all the points in $R(I,j+1)$, $R(I,j+2)$ and $R(I,j+3)$.
\begin{algorithm}[t]
	\DontPrintSemicolon
	\SetAlgoLined
	Vehicle stays at origin from time 0 (time first intruder arrives) to time $2\rho$, same side is right of origin\\
	At time $2\rho$, %$\max\{\frac{1-\rho}{v}-4\rho,2\rho\}$\\
	\eIf{$\abs{\subscr{S}{opp}^{1}} > \abs{\subscr{S}{same}^{1}}$}
	{Move to $x_{2} = -\rho$, same side is left of origin \;}
	{Move to $x_{2} = \rho$, same side unchanged}
    Wait until time $z$\\
	\For{each time instant $z+2\rho j, j\geq 0$}{
	\eIf{$\abs{\subscr{S}{opp}^{j+2}} > \abs{\subscr{S}{same}^{j+1}} + \abs{\subscr{S}{same}^{j+2}} + \abs{\subscr{S}{same}^{j+3}}$}{
		Move to $x_{j+1} = -x_j$, same side changes \;} % and capture interval $\subscr{S}{opp}^{j+2}$ \;}
	{ Stay at $x_j$ and capture interval $\subscr{S}{same}^{j+1}$.\;
	}}
	\caption{Capture with Patience Algorithm}
	\label{algo:CAP_IncP}
\end{algorithm}
First, we observe that the intruders in $\subscr{S}{same}^{j+1}$ are located between $\rho$ and $\rho + 2\rho v$, the intruders in $\subscr{S}{same}^{j+2}$ are located between $\rho + 2\rho v$ and $\rho+4\rho v$, and the intruders in $\subscr{S}{same}^{j+3}$ are located between $\rho+4\rho v$ and $\rho + 6\rho v$ (Fig. \ref{fig:CAP_region}).
Further, because $z \ge 6\rho$, this means all the intruders in $\subscr{S}{same}^{j+3}$ have arrived by time $2j \rho + z$. Similar conclusions can be drawn for the intruders in $\subscr{S}{opp}^{j+1}$, $\subscr{S}{opp}^{j+2}$, and $\subscr{S}{opp}^{j+3}$. 
If $\abs{\subscr{S}{opp}^{j+2}} > \abs{\subscr{S}{same}^{j+1}} + \abs{\subscr{S}{same}^{j+2}} + \abs{\subscr{S}{same}^{j+3}}$, then the vehicle moves to $-\rho$ arriving at time $2(j+1)\rho + z$ which is just in time to capture all the intruders in $\subscr{S}{opp}^{j+2}$. 
If not, then the vehicle stays at $\rho$ and captures all the intruders in $\subscr{S}{same}^{j+1}$ and reevaluates at time $2(j+1)\rho +z$.
The key observation is that the vehicle moves from $\rho$ to $-\rho$ only when it sees sufficient benefit in terms of the number of intruders in $\subscr{S}{opp}^{j+2}$ to sacrifice all the intruders in $\subscr{S}{same}^{j+1}$, $\subscr{S}{same}^{j+2}$ and $\subscr{S}{same}^{j+3}$.

% For the initial case at time 0, the vehicle stays at the origin until time $4\rho$. At this time, the first two intervals of points have arrived.
% If $|L(I,1)| > |R(I,1)| + |R(I,2)|$, then the vehicle will move to $-\rho$.
% If $|R(I,1)| > |L(I,1)| + |L(I,2)|$, then the vehicle will move to $\rho$.
% Otherwise, if $|L(I,1)| \ge |R(I,1)|$, then the vehicle will move to $-\rho$.
% Otherwise, the vehicle moves to $\rho$.
% Assume without loss of generality the vehicle is at position $\rho$ at time $5\rho$.
% The vehicle remains there until time $z$.
% At time, $z$, if $|L(I,2)| > |R(I,1)| + |R(I,2)| + |R(I,3)|$, then the vehicle moves to $-\rho$ arriving at time $2 \rho + z$ which is just in time to capture all the points in $L(I,2)$. 
% Otherwise, it stays at $\rho$ and captures all the points in $R(I,1)$ and then reevaluates at time $z+2\rho$.
For the initial case, the vehicle stays at the origin until time $2\rho$. %For simplicity, we define the time the first intruder arrives to be time 0.
At time $2\rho$,
%at time 0, the vehicle stays at the origin until time $\max\{\frac{1-\rho}{v}-4\rho,2\rho\}$ after the first intruder has arrived in the environment. %At this time, the first intervals of intruders have arrived.
%If $\abs{\subscr{S}{L}^{1}} > \abs{\subscr{S}{R}^{1}} + \abs{\subscr{S}{R}^{2}}$, then the vehicle will move to $-\rho$.
%If $\abs{\subscr{S}{R}^{1}} > \abs{\subscr{S}{L}^{1}} + \abs{\subscr{S}{L}^{2}}$, then the vehicle will move to $\rho$.
if $\abs{\subscr{S}{opp}^{1}} > \abs{\subscr{S}{same}^{1}}$ (intruders to the left of the origin are considered on the opposite side while intruders to the right of the origin are considered on the same side for this special case), then the vehicle will move to $-\rho$.
Otherwise, the vehicle moves to $\rho$.
In either case, the vehicle then stays at either $-\rho$ or $\rho$ until time $z$. 

\vspace{-0.1in}
\begin{lemma}
Algorithm \ref{algo:CAP_IncP} never moves the vehicle from $\rho$ to $-\rho$ and then back to $\rho$ (or vice versa) without capturing at least one interval of intruders.
\label{lem:pingpong}
\end{lemma}
\begin{proof}
This holds as in order to move from $\rho$ to $-\rho$ at time $2j\rho +z$, it must be true that $\abs{\subscr{S}{opp}^{j+2}} > \abs{\subscr{S}{same}^{j+1}} + \abs{\subscr{S}{same}^{j+2}} + \abs{\subscr{S}{same}^{j+3}}$.
This implies $\abs{\subscr{S}{opp}^{j+2}} > \abs{\subscr{S}{same}^{j+3}}$.
In order to move directly to $\rho$ without capturing the intruders in the interval $\subscr{S}{opp}^{j+2}$, we would need $\abs{\subscr{S}{same}^{j+3}} > \abs{\subscr{S}{opp}^{j+2}} + \abs{\subscr{S}{opp}^{j+3}} + \abs{\subscr{S}{opp}^{j+4}}$, but this cannot hold by the above observation.
\end{proof}

\begin{corollary}
For any $i \ge 1$, Algorithm \ref{algo:CAP_IncP} will capture intruders from one of $\subscr{S}{same}^{i}$ or $\subscr{S}{opp}^{i+1}$.
\label{cor:CAP-pattern}
\end{corollary}
\begin{proof}
At time $z + 2\rho(i-1)$, the vehicle is in position to capture $\subscr{S}{same}^{i}$ by definition of the algorithm. If it captures, $\subscr{S}{same}^{i}$, we are done. If not, it moves to the opposite capture point and by Lemma \ref{lem:pingpong} will capture $\subscr{S}{opp}^{i+1}$.
\end{proof}

\begin{theorem}\label{thm:4_comp_IncP}
Algorithm \ref{algo:CAP_IncP} is 4-competitive for any environment $\E$ with $v \le (1-\rho)/6\rho$.
\end{theorem}

\begin{proof}
The basic idea is that for any input instance $I$, Algorithm \ref{algo:CAP_IncP} captures at least $1/4$ of all intruders in $I$. We prove this claim using an accounting analysis where we ``charge'' lost intruders to  captured intruders, { or equivalently, captured intruders ``pay'' for the lost intruders}.
One notation note: our proof will first focus on a captured interval which we identify as being on the same side and then focus on a lost interval which we identify as being on the opposite side. All other intervals will be defined as being on the same side or opposite side relative to these anchor intervals.

We first describe how intervals of captured intruders $\subscr{S}{same}^{j}$ pay for intervals of lost intruders.
We divide intervals of captured intruders into two types:
type (a) where the vehicle did not move to capture them meaning the vehicle also captured the previous interval on the same side (or the interval corresponds to the first interval $\subscr{S}{same}^{1}$) and 
type (b) where the vehicle did move to capture them meaning the vehicle spent the last $2\rho$ time moving from $\rho$ to $-\rho$ or vice versa.
A type (a) captured interval $\subscr{S}{same}^{j}$ will be charged three times, once each to pay for each of the lost intervals $\subscr{S}{opp}^{j-1}$, $\subscr{S}{opp}^{j}$, and $\subscr{S}{opp}^{j+1}$.
A type (b) captured interval $\subscr{S}{same}^{j}$ will also be charged three times, one charge to pay all at once for lost intervals $\subscr{S}{opp}^{j-1}$, $\subscr{S}{opp}^{j}$, and $\subscr{S}{opp}^{j+1}$,
one charge to help pay for lost interval $\subscr{S}{same}^{j-1}$, and one charge to help pay for lost interval $\subscr{S}{same}^{j-2}$.
Because each captured interval is only charged three times, if we can show that every lost interval of intruders is fully paid for by this charging scheme, we will have proven that each captured intruder pays for at most three lost intruders and the result follows.

We now show that every interval of lost intruders $\subscr{S}{opp}^{j}$ is fully paid for by this charging scheme.
We divide intervals of lost intruders into five types:
type (a) where the intervals $\subscr{S}{same}^{j-1}$, $\subscr{S}{same}^{j}$, and $\subscr{S}{same}^{j+1}$ are all captured,
type (b) where the intervals $\subscr{S}{opp}^{j-2}$ and $\subscr{S}{same}^{j}$ are captured, 
type (c) where the intervals $\subscr{S}{opp}^{j-1}$ and $\subscr{S}{same}^{j+1}$ are captured,
type (d) where the intervals $\subscr{S}{same}^{j-1}$ and $\subscr{S}{opp}^{j+1}$ are captured, and
type (e) where the intervals $\subscr{S}{same}^{j-1}$, $\subscr{S}{same}^{j}$, and $\subscr{S}{opp}^{j+2}$ are captured.

We first claim these five types of intervals describe all possible variations of lost intervals ignoring boundary cases (the first time interval and the last time interval).
This follows from Corollary \ref{cor:CAP-pattern} which shows the algorithm will never fail to capture intruders from both sides from two consecutive $2\rho$ time intervals.
Type (a) lost intervals are fully paid for by the captured intervals $\subscr{S}{same}^{j-1}$, $\subscr{S}{same}^{j}$, and $\subscr{S}{same}^{j+1}$ because the algorithm at time $z+2\rho(j-2)$ did not choose to switch sides and capture $\subscr{S}{opp}^{j}$, so $\abs{\subscr{S}{opp}^{j}} \le \abs{\subscr{S}{same}^{j-1}} + \abs{\subscr{S}{same}^{j}} + \abs{\subscr{S}{same}^{j+1}}$.

The argument for type (b) and type (c) lost intervals are essentially identical to each other.
Type (b) lost intervals are fully paid for by the captured interval $\subscr{S}{same}^{j}$ because the algorithm at time $z+2\rho(j-2)$ did choose to switch sides to capture $\subscr{S}{same}^{j}$ which means that $\abs{\subscr{S}{same}^{j}} > \abs{\subscr{S}{opp}^{j-1}} + \abs{\subscr{S}{opp}^{j}} + \abs{\subscr{S}{opp}^{j+1}}$.
Type (c) lost intervals are fully paid for by the captured interval 
$\subscr{S}{same}^{j+1}$ because the algorithm at time $z+2\rho(j-1)$ did choose to switch sides to capture $\subscr{S}{same}^{j+1}$ which means that $\abs{\subscr{S}{same}^{j+1}} > \abs{\subscr{S}{opp}^{j}} + \abs{\subscr{S}{opp}^{j+1}} + \abs{\subscr{S}{opp}^{j+2}}$. 

Next, the argument for type (d) and type (e) lost intervals are essentially identical to each other.
Type (d) lost intervals are fully paid for by the captured intervals $\subscr{S}{same}^{j-1}$ and $\subscr{S}{opp}^{j+1}$ for the following reasons.
First, because the algorithm at time $z+2\rho(j-2)$ did not choose to switch sides and capture $\subscr{S}{opp}^{j}$, it follows that $\abs{\subscr{S}{opp}^{j}} \le \abs{\subscr{S}{same}^{j-1}} + \abs{\subscr{S}{same}^{j}} + \abs{\subscr{S}{same}^{j+1}}$. Second, because the algorithm at time $z+2\rho(j-1)$ did choose to switch sides to capture $\subscr{S}{opp}^{j+1}$,  $\abs{\subscr{S}{opp}^{j+1}} > \abs{\subscr{S}{same}^{j}} + \abs{\subscr{S}{same}^{j+1}} + \abs{\subscr{S}{same}^{j+2}}$. The full payment follows from combining these two observations.
The argument for type (e) lost intervals is essentially the same except they are paid for by the captured intervals $\subscr{S}{same}^{j-1}$, $\subscr{S}{same}^{j}$, and $\subscr{S}{opp}^{j+2}$.

Finally the boundary cases of the first intervals and the last intervals fall into these five types if we add dummy intervals  $\subscr{S}{same}^{0}$ and $\subscr{S}{same}^{Y+1}$, all with cardinality 0, where $Y$ denotes the last interval with actual intruders on either side. We assume the vehicle captures both of these dummy intervals.

Since we have shown our charging scheme does pay for all lost intruders and each captured intruder pays for at most three lost intruders, the result follows.
\end{proof}

We now prove some lower bounds on the competitive ratio for the CAP algorithm including showing that the bound is tight for some parameter settings. 

\vspace{-0.2in}
\begin{lemma}
    Algorithm \ref{algo:CAP_IncP} is no better than 3-competitive for $v\leq \frac{1-\rho}{6\rho}$ and is no better than 4-competitive for $v\leq \min\{\frac{1}{3},\frac{1-\rho}{6\rho}\}$.
\end{lemma}
\begin{proof}
We prove these results using an input instance $I$ consisting of two streams of intruders, one stream each arriving at endpoints $1$ and $-1$. At location $1$, one intruder arrives at time instant $6i\rho$, $0 \le i \le K$ for some very large $K$. At location $-1$, one intruder arrives at time instant $3\rho+\frac{\rho}{v}+i2\rho$, %$2\rho<t<4\rho$ and 
$0 \le i \le K$ for the same $K$.

As the first intruder at location $1$ arrives $3\rho+\frac{\rho}{v}$ time units before the first intruder arrives at location $-1$, the CAP algorithm moves the vehicle to $\rho$ from the origin, reaching $\rho$ at time $3\rho$.
% (Fig. \ref{fig:example_CAP} (a)). 
%This is because until time instant $1+\max\{\frac{1-\rho}{v}-4\rho,2\rho\}$, there will be $a$ intruders in $\subscr{S}{same}^1$ and none in $\subscr{S}{opp}^1$. 
As the intruders at location $1$ arrive every $6\rho$ time units and the intruders at location $-1$ arrive every $2\rho$ time units, from this moment on until the end, the vehicle remains at $\rho$ because there will always be exactly one intruder in $\subscr{S}{opp}^{j+2}$ and exactly one intruder in $\subscr{S}{same}^{j+1}$, $\subscr{S}{same}^{j+2}$ and $\subscr{S}{same}^{j+3}$ combined. 
% (Fig. \ref{fig:example_CAP} (b)). 
%After time instant $1+\frac{1-\rho}{v}$, the vehicle captures the first $a$ intruders. After time instant $1+\frac{1-\rho}{v}$, there will always be $a$ intruders in $\subscr{S}{opp}^{j+2}$ and $a$ intruders in $\subscr{S}{same}^{j+1}$, $\subscr{S}{opp}^{j+2}$ and $\subscr{S}{opp}^{j+3}$ combined (Fig. \ref{fig:example_CAP} (b)). 
Thus, except at the beginning and at the end of the streams, every $6\rho$ time units, the vehicle captures $1$ intruder that arrives at $+1$ but loses $3$ intruders that arrive at $-1$.

For $v\leq \frac{1-\rho}{6\rho}$, there is an obvious offline algorithm that captures $3/4$ of the intruders; namely, one that moves the vehicle to $-\rho$ and captures all the intruders that arrive at location $-1$. Thus, for the specified parameter settings, the CAP algorithm cannot be better than $3$-competitive.

We now show that the optimal algorithm can capture all intruders for specific parameter settings. First consider $\rho \leq \frac{1}{2}$. Let $p_1$ and $p_2$ denote two distinct points which are $\rho$ and $-2\rho$ distance away from the origin respectively.
% and $p_2$ denote a point which is $-2\rho$ distance away from the origin.

The optimal algorithm moves the vehicle towards $p_1$ at time instant $\frac{1-\rho}{v}-\rho$ reaching $p_1$ just in time to capture the first intruder that arrived at endpoint $1$. Since the first intruder at endpoint $-1$ arrives $3\rho+\frac{\rho}{v}$ time after the arrival of first intruder at endpoint $+1$, the vehicle captures the second intruder at $p_2$ by moving immediately towards $p_2$. This is because the intruder will be located at a distance of $3\rho v+\rho$ from $-\rho$ at the time when the vehicle captured the intruder at $p_1$. This means that the distance between the vehicle and the intruder will be $3\rho+3\rho v$ implying that the vehicle captures this intruder in $3\rho$ time or equivalently at location $p_2$.
% \erictodo{Maybe expand above sentence to show this more explicitly}
Note that while moving from $p_1$ to $p_2$ and back to $p_1$ the vehicle takes $6\rho$ time units. Thus, the vehicle can capture every intruder, that arrives at endpoint $1$ at time instant $6i\rho$, at location $p_1$ at time instant $\frac{1-\rho}{v}+6i\rho$. 
%\erictodo{I think the exact time should be $6i(1-\rho)/v$ since it may be the case it takes the intruder more than 6$\rho$ time to get to $\rho$.}\textcolor{ForestGreen}{I think the exact time is $\frac{1-\rho}{v}+6i\rho$ because the first intruder will be captured in $\frac{1-\rho}{v}$ time and every $6\rho$ time after that since intruders arrive after every $6\rho$.}

Now let us consider the intruders that arrive at $-1$.
Note that the intruders that arrive at $-1$ arrive $2\rho$ time units apart and we already showed that the first intruder that arrived at $-1$ can be captured at $p_2$. This means that from the moment the vehicle leaves $p_2$ after capturing an intruder, the next intruder will take $2\rho$ time to reach $p_2$ and $2\rho+\frac{\rho}{v}$ time units to reach $-\rho$, whereas the vehicle will take $5\rho$ time to move from $p_2$ to $p_1$ and then to $-\rho$. Thus, in order to ensure that the next intruder is not lost while the vehicle moves from $p_2$ to $p_1$ and then to $-\rho$, we need $5\rho\leq 2\rho+\frac{\rho}{v}$ implying $v\leq\frac{1}{3}$. In summary, after capturing an intruder at $p_1$, the vehicle moves to $p_2$, capturing an intruder that arrives at $-1$ at time $3\rho+\frac{\rho}{v}+6i\rho$ at location $p_2$. Since $v\leq \frac{1}{3}$, the intruders that arrive at times $5\rho+\frac{\rho}{v}+6i\rho$ and $7\rho+\frac{\rho}{v}+6i\rho$ are captured on the way to $p_2$.

We now consider the case when $\rho>\frac{1}{2}$. Since $\rho>\frac{1}{2}$, we cannot set the point $p_2$ at a distance of $-2\rho$; instead, we set $p_2$ to be $-1$ and have the vehicle idle at $-1$ for $2(2\rho-1)$ time. 
Note that the total time that the vehicle takes to move from $p_1$ to $-1$ combined with the waiting time at $-1$ is $5\rho-1$.
%\erictodo{Is this time no $6\rho$ instead of $5\rho-1$?}\textcolor{ForestGreen}{ I believe not as this is the time the vehicle takes to move from $p_1$ to -1, which is $1+\rho$, combined with the total time the vehicle stays idle at $-1$, i.e., $2(2\rho-1)$. So the total will be $1+\rho+2(2\rho-1)=5\rho-1$. $6\rho$ will be the time the vehicle takes when it moves from $p_1$ to -1, stays idle at $-1$ then moves back to $p_1$, which is $5\rho-1+1+\rho=6\rho$.}
This time is sufficient for the vehicle to capture the intruder that arrives at time $3\rho+\frac{\rho}{v}+6i\rho$ immediately. The next intruder, however, arrives at time $3\rho+\frac{\rho}{v}+6i\rho$ which is after the vehicle leaves $-1$ as the vehicle leaves $-1$ at time $5\rho-1+6i\rho$. This is equivalent to the intruder arriving at $-1$, $1+\frac{\rho}{v}$ time after the vehicle has left $-1$.
The total time the vehicle takes to move from $-1$ to $\rho$ and then to $-\rho$ is $1+3\rho$ and the intruder will take $1+\frac{\rho}{v}+\frac{1-\rho}{v}$ time.
Thus, in order to ensure that the next intruder is not lost, we need $3\rho\leq \frac{1}{v}$ implying $v\leq \frac{1}{3\rho}$. Since, $v\leq \frac{1}{3}$ from the previous case, $v\leq \frac{1}{3\rho}$ always holds. Furthermore, since the algorithm is defined for $v\leq \frac{1-\rho}{6\rho}$, we get the result.
\end{proof}

% \sdb{[Where is the condition that $v \leq (1-\rho)/(6\rho)$ being used? Is it coming from the analysis in Theorem~\ref{thm:4_comp_IncP}? If so, it must be specified.]}
 
% The optimal algorithm starts at the origin and moves the vehicle towards $1$ at time instant $0$ capturing the first intruder at endpoint $1$ at time instant $1$. The vehicle then moves at maximum speed towards $-1$ to capture the first intruder that arrives at $-1$ before it reaches $-\rho$. This first intruder arrives at time $1+3\rho$, so we need $1+\rho \leq \frac{1-\rho}{v}+3\rho$ which holds since $v\leq 1$. 

% \erictodo{Some cases here need work. First, we cannot throw away the bound of $v \le \frac{1-\rho}{3-5\rho}$. This might be tighter than $v \le \frac{1-\rho}{6\rho}$. Second, we cannot assume that the vehicle will stop and capture later intruders at $\rho$. This requires synchronization and a specific $v$. }

% Now, in order to ensure that the vehicle captures all the intruders in $[-1,-\rho]$ and the intruders that arrived at the endpoint $1$ at time instant $1+6\rho$, we require $3+\rho\leq \frac{(1-\rho)}{v}+6\rho$ implying $v\leq \frac{1-\rho}{3-5\rho}$. Finally, for $i>1$, the vehicle will be located at $\rho$ and can move from $\rho$ to $-1$ and back to $\rho$ just in time to capture the intruders that arrives at time instant $1+6i\rho$. Thus, $\frac{2(1+\rho)}{1+v}\leq 6\rho$, yielding $v\geq\frac{1-2\rho}{3\rho}$. Since the algorithm is defined for $v\leq \frac{1-\rho}{6\rho}$, we get the result.

In the CAP algorithm, the vehicle waits for intruders at $\rho$ or $-\rho$ to capture them. Furthermore, CAP is \emph{memoryless}, i.e., it depends only the present state of the vehicle and the intruders.
% By doing so, we established that the CAP algorithm is $4$-competitive. 
In the following subsection, we formulate another memoryless algorithm that moves the vehicle beyond the perimeter and analyze its performance.

\begin{figure}[t]
    \centering
    \includegraphics[scale=0.45]{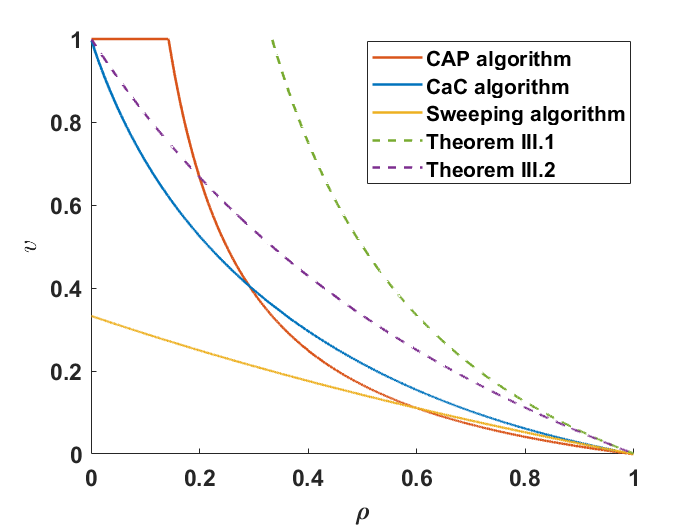}
    \caption{Parameter regimes for our algorithms (solid lines, extend to the left) and lower bounds (dashed lines, extend to the right).
    }
    \label{fig:v_rho_curve}
    \vspace{-0.6cm}
\end{figure}

\section{Summary and Numerical Performance}\label{sec:numerics}

\subsection{Summary of the results}
Figure \ref{fig:v_rho_curve} shows a $v$-$\rho$ plot summarizing our results. 
% Given a value of $\rho$, all values of speed $v$ that lie below a curve satisfy the conditions for the competitiveness of the respective algorithm. In the region above the curve defined by Theorem \ref{thm:Inc_no_c}, there is no $c$-competitive algorithms for any $c$.
% The following observations are in order:
% \begin{enumerate}
%     \item 
For $\rho >0.3$, as the curve defined by the conditions in Theorem \ref{thm:2_comp} for the CaC algorithm is above the curve defined by the conditions in Theorem~\ref{thm:4_comp_IncP} for the CAP algorithm, one should always implement the CaC algorithm rather than the CAP algorithm.
    % \item 
For any $\rho \leq 0.3$, there exist values of $v$ such that one might choose any of the three algorithms.
    % \item 
The curve defined by the conditions in Theorem \ref{thm:2_comp} for the CaC algorithm is completely below the curve defined by the condition in Theorem \ref{lem:bound_2_comp_IncP}. This suggests that for the values of $v$ and $\rho$ that lie above the curve defined by the conditions for the CaC algorithm, either there may still exist an algorithm which is 2-competitive or it may be possible to tighten the analysis of  Theorem~\ref{lem:bound_2_comp_IncP} and Lemma~\ref{lem:condition}.
% \end{enumerate} 

\subsection{Numerical Performance}
We now analyze the average case performance, as opposed to the worst case performance, of our algorithms numerically.
Of particular interest is the case when the intruders are generated stochastically as per a spatio-temporal arrival process \cite{bertsimas1991stochastic}. 

We performed numerical analysis of our algorithms using the following procedure.
A Poisson process with rate $\lambda$ was used to model the arrival process of the intruders. The intruders arrive with equal probability on the endpoints.
We simulated 50 runs per algorithm and present the mean and standard deviation of the \emph{capture fraction} obtained by each algorithm for various values of $v$ keeping $\rho$ and $\lambda$ fixed to $0.2$ and $5$ respectively. The \emph{capture fraction} is defined as the ratio of the total number of intruders captured to the total number of intruders arrived in the environment \cite{Smith2009translating}. The value of $\lambda$ was kept high because for low arrival rate ($\lambda \to 0$), the number of intruders that arrive in the environment were very few and the capture fraction obtained was misleadingly high. 
% For values of $\lambda>1$, the capture fraction obtained by the algorithms was constant on average for fixed values of $v$ and $\rho$.

\begin{figure}[t]
    \centering
    \includegraphics[scale=0.42]{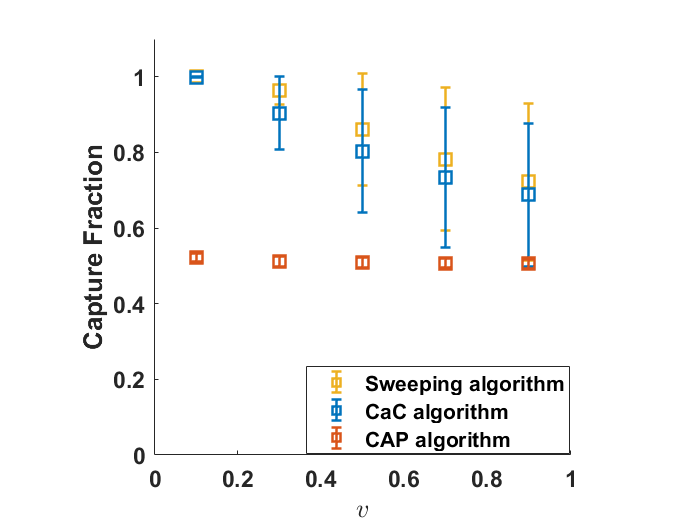}
    \caption{Simulation result for $\lambda=5$ and $\rho=0.2$. The error bars indicate $\pm 1$ standard deviation.}
    \label{fig:IncP_num}
    \vspace{-0.6cm}
\end{figure}

Figure \ref{fig:IncP_num} shows the simulation result for each of the algorithms. For values of $v$ and $\rho$ that lie above the blue curve in Figure \ref{fig:v_rho_curve}, the CaC algorithm captured, on average, more than half of the intruders that arrived. Furthermore, the capture fraction of CAP algorithm was approximately $0.5$ on average for all values of $v$. This is because the intruders being uniformly distributed, the vehicle can just capture intruders on one side and still ensure at least half of the total intruders are captured. Moreover, as $v$ increases the capture fraction of Sweep algorithm approaches that of CaC algorithm. This is because as $v$ increases, the sizes of the sets $\subscr{S}{same}$ and $\subscr{S}{opp}$ increase and eventually they cover the entire environment, thereby, converging to the Sweep algorithm.

\section{Conclusion and Future Work}\label{sec:conclusion}

This paper addressed a problem in which a single vehicle is tasked to defend a line segment perimeter from  intruders.
The key novelty of this work is an integration of concepts and techniques from competitive analysis of online algorithms with pursuit of multiple mobile intruders. We designed and analyzed three algorithms, i.e., Sweeping, Compare and Capture, and Capture with Patience algorithms, and demonstrated that they are $1$, $2$ and $4$-competitive, respectively. We also derived fundamental limits on $c$-competitiveness for any constant $c$.

We plan to extend this work for the case when the intruders need to move outward or can actively evade the vehicle in order to reach the perimeter.
Cooperative multi-vehicle defense in higher dimensional environments that can yield lower competitive ratios is another future direction.

\bibliographystyle{IEEEtran}
\bibliography{reference}

\end{document}